\documentclass[prl, twocolumn, showpacs]{revtex4}

\begin{document}

\title{Reply to Comment on "Modified Coulomb Law in a Strongly Magnetized Vacuum"
}

\author{A.E. Shabad}

 \email{shabad@lpi.ru}
\affiliation{%
P.N. Lebedev Physics Institure, Moscow, Russia\\}

\author{V.V. Usov}
 \email{fnusov@wicc.weizmann.ac.il}
\affiliation{Center for Astrophysics, Weizmann Institute of Science,
Rehovot 76100, Israel\\}


\begin{abstract} This is a reply to the Comment by S.-Y. Wang
 concerning our paper "Modified Coulomb Law in a
Strongly Magnetized Vacuum"
\end{abstract} 

 \maketitle

 \noindent {\bf Shabad and Usov Reply:} In the commented
paper \cite{PRL07} (see also our more detailed subsequent
publication \cite{arXiv07}) our purpose was to study the $static$
interaction between electric charges placed in a strong magnetic
field by considering the effect of vacuum polarization, modifying
the Coulomb potential of a point charge. It is understood that any
use of a $static$ potential for considering an atomic spectrum is
$apriory$ confined exclusively to a nonrelativistic approximation,
with
retardation and recoil effects 
disregarded. For this reason we did not
proceed beyond the Schr$\ddot{\rm o}$finger equation with the
static potential (although a consideration of the Dirac and
Bethe-Salpeter (B.S.) equation with equal-time Ansatz
might be also reasonable - see a discussion in \cite{arXiv07}).
 We found that the vacuum polarization
prevents the known trend of the Schr$\ddot{\rm o}$dinger ground
state energy to negative infinity as the magnetic field infinitely
grows. Magnetic fields that may be referred to as asymptotically
large in our context are a few orders of magnitude larger than
$(2\pi /\alpha) B_0 \sim 10^3 B_0$, $B_0=m^2/e$.
An essential deviation of the ground state
energy from the widely recognized result, which tends to
$-\infty$, is achieved already for the fields of the order of
$10^4B_0$ or $10^5B_0$, i.e. for those that may be of interest in
pulsar physics, whereas for the field $\sim 10^{11}B_0$ all the
limits are already saturated within reasonable accuracy, formation
of the delta-function and the string potential \cite{arXiv07} is
completed, in other words such field is the physical infinity
within the scale of the problem. Magnetic fields, for which,
according to the author of Comment \cite{wang}, $m_{\rm dyn}$
becomes of the order of the electron mass $m$ or more, is,
following his Eq. (1), many orders of magnitude larger,
$B>\alpha^{-2}\exp (2\pi/\alpha\ln\alpha^{-1})B_0\cong 2\times
10^{80}B_0$.
It is true that to magnetic fields of such
a scale our results stated in \cite{PRL07}, \cite{arXiv07}
are not and should not be applicable, because of the need of a fully
relativistic treatment.

In the series of papers quoted in  Comment \cite{wang} the
appearance of dynamical mass of a primarily massless electron,
called magnetic catalysis, was achieved thanks to the presence of
magnetic field that supplies the necessary dimensional parameter
and thus destroys the initial chiral invariance peculiar to the
massless QED.  For instance, in the most recent work \cite{leung}
in this series, 
a nonperturbative mechanism has led to the dynamical mass fitted
as Eq. (1) 
via
  solution of the Schwinger-Dyson set, truncated in a certain way.
To keep the dynamical mass $m_{\rm dyn}$ equal
to the experimental mass of electron $m$ the above tremendous
magnetic fields would be necessary. This is a comprehensible
consequence of the fact that Eq. (1) in \cite{wang} relates to the
massless electrodynamics, the free electron propagator without
mass being laid into the underlying  equations. Among the results
of Ref. \cite{leung}, we do not see a sufficient basis for the
proposition made in Comment that "in $realistic$ QED with massive
electrons, the electron in an extremely strong magnetic
field will aquire a mass" with "the asymptotic behavior
$m_\ast\approx m_\mathrm{dyn}\to\infty$ as $B\to\infty$",
i.e. the same as in massless QED.

In our previous publications \cite{prl} we did consider the
interaction between charged particles for the huge fields,
referred to in Comment, in a fully relativistic way, namely the
B.S. equation for a mutually bound electron-positron
pair. Before the radiative corrections
are taken into account this problem has the same
feature as the Schr$\ddot{\rm o}$dinger equation: the ground state
level for the positronium atom sinks unlimitedly and is able to
reach the zero-energy value like it occurs in an atom with a
supercharged nucleus with $Z>\alpha^{-1}.$ This happens for
magnetic fields of the order of $B\simeq
\frac{B_0}{4}\exp\{\frac{\pi^{3/2}}{\sqrt{\alpha}}\}\sim
10^{24}B_0.$ With this magnetic field the energy gap between the
electron and positron shrinks and the vacuum becomes unstable.
This puts an upper limit on magnetic fields admitted in QED.
Therefore, the indicated values of 10$^{80}B_0$ cannot be
achieved. This conclusion of Ref. \cite{prl} may be invalidated,
indeed, if S.-Y. Wang succeeds to demonstrate the validity of his
statement quoted above, because in this case the growing effective
 electron mass $m_\ast$ when substituted inside the Dirac propagator
 in the B.S. equation would prevent the gap from shrinking.

A different circumstance is notable for the present discussion:
the sinking of the fully relativistic ground level is not stopped
 by the vacuum
polarization \cite{prl}, nor by the perturbative corrections to the electron
propagator. The reason lies in  that all the three photon modes
provide the infinite deepening of the level, since they all
contribute into the pole $1/x^2$ of the photon propagator on the
light cone $x^2=0$ , whereas only the mode 2, responsible in
particular for electrostatic force, possesses a part linearly
growing with the magnetic field that causes suppression of its
contribution into the interaction between charged particles. This
differs from our nonrelativistic result in \cite{PRL07} without
contradicting it, since it relates to much larger fields that make
the relativistic approach nesessary. In other words, within the relativism
the nonstatic photon modes support the infinite deepening of the level,
suppressed in nonrelativism by the static mode -
the only one in that case.

Supported by RFBR project  05-02-17217 and  
Programme LSS-4401.2006.2, and by the Israel Science Foundation of
IASH.





\medskip

\noindent Received $\;\;$ September 25, 2007; 



\end{document}